\documentclass{an}
\usepackage{graphicx}
\usepackage{times}
\usepackage{fancyhdr}
\newcommand{\Msolar}{\mbox{$M_{\odot}\,$}}
\newcommand{\Lsolar}{\mbox{$L_{\odot}\,$}}
\def\gs{\mathrel{\raise0.35ex\hbox{$\scriptstyle >$}\kern-0.6em \lower0.40ex\hbox{{$\scriptstyle \sim$}}}}
\def\ls{\mathrel{\raise0.35ex\hbox{$\scriptstyle <$}\kern-0.6em \lower0.40ex\hbox{{$\scriptstyle \sim$}}}}
\newcommand{\arcsecs}{\mbox{$^{\prime\prime}$}}

\begin{document}

\title{350-$\mu$m {\sc sharc-ii} Imaging of Luminous High-$z$ Radio Galaxies}

\author{T.R. Greve\inst{1}
\and  R.J. Ivison\inst{2,3}
\and J.A. Stevens\inst{4}}
\institute{
California Institute of Technology, Pasadena, CA 91125, USA
\and 
Astronomy Technology Centre, Royal Observatory, Blackford Hill, Edinburgh EH9 3HJ, UK
\and
Institute for Astronomy, University of Edinburgh, Blackford Hill, Edinburgh EH9 3HJ, UK
\and 
Centre for Astrophysics Research, Science and Technology Research Centre, 
University of Hertfordshire, College Lane, Herts AL10 9AB, UK}

\date{Received $<$date$>$; 
accepted $<$date$>$;
published online $<$date$>$}

\abstract{Using the filled bolometer array {\sc sharc-ii} on the Caltech Submillimeter
Observatory (CSO), we have obtained deep ($\sigma_{rms}\sim 15$\,mJy\,beam$^{-1}$), 
high-quality 350-$\mu$m maps of five of the most
luminous high-$z$ radio galaxies known. In all cases the central radio galaxy is
detected at the $\gs 3\sigma$ level, and in some cases the high resolution 
of {\sc sharc-ii} ({\sc fwhm}~$=9\arcsecs$) allows us to confirm the
spatially extended nature of the dust emission.
In PKS\,1138$-$262 ($z=2.156$), 8C\,1909$+$722 ($z=3.538$) and
4C\,41.17 ($z=3.792$), additional sources -- first discovered
by {\sc scuba} at 850-$\mu$m and believed to be dusty, merging systems associated with
the central radio galaxy -- are detected at 350-$\mu$m. Furthermore, in PKS\,1138 and
4C\,41.17 additional {\sc sharc-ii} sources are seen which were not detected at 850-$\mu$m,
although the reality of these sources will have to be confirmed by independent submm observations.
Thus, our observations seem to support the notion of extended star formation taking place
in radio galaxies at high redshifts, and that these systems 
sit at the centers of overdense regions harbouring a complexity of dusty and 
vigorously star forming systems. At the redshift of the radio galaxies,
the 350-$\mu$m observations sample very close to the rest-frame dust peak (typically
at $\sim 100$-$\mu$m), and they therefore
contribute a particularly important point to the spectral energy distributions of these sources, which
we use this in conjunction with existing (sub)millimeter data to derive FIR luminosities, 
dust temperatures and spectral emissivities of the central radio galaxies.
}

\correspondence{tgreve@submm.caltech.edu}

\maketitle

\section{Introduction}
High-$z$ Radio Galaxies (HzRGs) are typically identified in low frequency radio surveys
as Ultra Steep Spectrum (USS) radio sources (De Breuck et al.\ 2000), which
in conjunction with near-IR colour criteria provide a highly
efficient selection scheme. 
The most powerful ($P_{178MHz}\gs 10^{28}\,h^{-2}$\,W\,Hz$^{-1}$)
HzRGs have become an invaluable tool 
for testing models of galaxy formation, as radio galaxies are amongst the
most massive galaxies at all redshifts, and as such they trace out the most
biased peak in the primordial density field.
Furthermore, owing to their accurate radio positions and morphologies, as well
as their extreme properties, such as large (optical) luminosities,
extended X-ray and Ly$\alpha$-halos, HzRGs are relatively
easy to study compared to other populations of high-$z$ galaxies.\\
\indent The first pointed submillimeter (submm) observations of high-$z$ radio galaxies demonstrated
that a large fraction of such systems are extremely FIR luminous ($L_{\mbox{\tiny{FIR}}}\sim 10^{13}\,\Lsolar$)
and contain large amounts of dust ($M_d\sim 10^9\,\Msolar$ 
 -- Dunlop et al.\ 1994; Chini \& Kr\"{u}gel 1994; Ivison et al.\ 1995).
Subsequently, systematic 850-$\mu$m surveys with SCUBA of HzRGs by Archibald et al.\ (2001)
and Reuland et al.\ (2004) confirmed these initial findings, and 
showed that the dust content of HzRGs increases strongly with redshift.
The latter was interpreted as HzRGs being increasingly younger (younger
stellar populations) at higher redshifts, since while the FIR luminosities 
can be powered by either Active Galactic Nuclei (AGN) or starburst activity, 
the large dust masses requires that vigorous star formation has recently been taking place.
Thus, even in HzRGs with their powerful AGNs, starburst activity is likely to contribute substantially to
the extreme FIR luminosities, albeit probably less so than in high-$z$ submm-selected galaxies
which are predominantly starburst dominated (Alexander et al.\ 2003).
Furthermore, about a handful of HzRGs are known to harbour large reservoirs of
molecular gas ($M(\mbox{H}_2)\sim 10^{11}\,\Msolar$ --  e.g.\
Papadopoulos et al.\ 2000; De Breuck et al.\ 2003; Greve et al.\ 2004;
De Breuck et al.\ 2005; Klamer et al.\ 2005) -- enough to fuel a $\sim 1000\,\Msolar\,$yr$^{-1}$
starburst for $\sim 10^8$\,yr. Finally, the peak of the dust and CO emission is
often not centered on the central radio galaxy, but offset and spatially extended
on scale of up to 30-200\,kpc.
Some of the most striking and visually gratifying examples of this came 
from submm (850-$\mu$m) SCUBA observations of seven high-$z$ radio galaxies,
which not only revealed extended ($\sim 30$\,kpc) dust emission from the central radio galaxy
itself, but also showed an overdensity by a factor of two compared to the field of sub-mm sources
around the radio galaxies (Stevens et.\ al 2003), suggestive of extended star formation
on hundreds of kilo-parsec scales.

\section{Results}
Here we present 350-$\mu$m lissajou-style scan-maps obtained with {\sc sharc-ii} (Dowell et al.\ 2002)
on the Caltech Submillimeter Telescope (CSO) on Hawaii of 5 of the 7 HzRGs
mapped with SCUBA at 850-$\mu$m by Stevens et al.\ (2003). 
The motivation for doing this was two-fold: 1) to utilise the
superior resolution of {\sc sharc-ii} ({\sc fwhm}~$=9\arcsecs$) over that of
{\sc scuba} ({\sc fwhm}~$=15\arcsecs$) to confirm the spatially extended dust emission
from the HzRGs, and the reality of the many companion objects reported by
Stevens et al.\ (2003), and 2) to determine accurate FIR luminosities and dust masses
in these objects. {\sc sharc-ii} observations are particular useful for the latter since
at the typical redshift of the HzRGs ($z\sim 3.5$) they probe the $\sim 100\,\mu$m rest-frame
peak of the dust emission.
Stevens et al.\ targeted these 7 sources because they were the brightest
sources at 850-$\mu$m of the SCUBA observed sample of HzRGs by Archibald et al.\ (2001).
Thus, it is important to keep in mind that our samples represents
the most luminous and extreme examples of radio galaxies.

\paragraph{PKS\,1138$-$262}
At 350-$\mu$m the central radio galaxy (abbreviated RG, and marked by a cross in Figure 1) 
is seen to be associated with
a $\sim 5\sigma$ emission feature consisting of two connected
knots roughly aligned with the axis defined by the kiloparsec scale radio jets (marked with
tick marks around the central RGs in Figure 1).
This double-knotted structure is most likely real since it is also discernable
at 850-$\mu$m, and in both maps the south-western most source 
($S_{350\mu m}=(92\pm 18)$\,mJy, $\sim 7\arcsecs$
south-west from the RG) is the brightest, see Table 1.
However, at the position of the RG, which coincided with the weaker
knot ($S_{350\mu m}=(61\pm 15)$\,mJy), a 1200-$\mu$m flux density of $(1.5\pm 0.5)$\,mJy 
was detected using {\sc mambo} on the IRAM 30\,m Telescope (Carilli et al.\ 2001).\\
\indent The brightest {\sc scuba} source ($S_{850\mu m}=(7.8\pm 2.2)$\,mJy) detected by Stevens et al.\ in this field
lies about 70\arcsecs east of the RG and coincides with a $1.9\sigma$ peak 
in the 350-$\mu$m map. Several explanations may account for the 
lack of a significant 350-$\mu$m detection of this very bright {\sc scuba} source:
a) the 850-$\mu$m detection could be spurious (it lies at the edge of the {\sc scuba} map),
b) the source lies near the edge of the {\sc sharc-ii} map where the noise
is higher which makes it less likely to be detectable, or c) the source is real, but at such a high redshift that the 350-$\mu$m band
is shifted beyond the dust peak and samples the SED so far down the Wien tail that it is undetectable
given the depth of the maps.
We are able to put a $3\sigma$ upper limit on its 350-$\mu$m flux of 
$S_{350\mu m} \ls 95$\,mJy.\\
\indent Another, much weaker {\sc scuba} source, $\sim 37\arcsecs$ west of the RG, 
coincide with  $\sim 2\sigma$ peaks in the 350-$\mu$m map. 
We derive an upper limit on its 350-$\mu$m flux of $S_{350\mu m}\ls 68$\,mJy.
Finally, a third {\sc scuba} source $\sim 57\arcsecs$ south of the RG 
falls outside the 350-$\mu$m map.
Thus, while we verify the double-knot structure of the central radio galaxy,
we do not confirm any of the apparently associated sources seen with {\sc scuba}.

\paragraph{8C\,1909$+$722}
The 350-$\mu$m {\sc sharc-ii} map shows a strong ($\gs 5\sigma$) 
central source ($S_{350\mu m}=90\pm 15$\,mJy), and a second significant ($\gs 4\sigma$) source 
about 38\arcsecs south-eastwards ($S_{350\mu m}\sim 148\pm 32$\,mJy). 
The former is offset from the RG-position by $\sim 6\arcsecs$, 
while the latter lies within $\sim 8\arcsecs$ of 
a strong {\sc scuba} source detected about 46\arcsecs south-east of the RG
(Stevens et al.\ 2003). These offsets can easily be accounted for considering 
that a) we have not corrected for any potential
astrometrical offsets between the {\sc sharc-ii} and {\sc scuba} maps, 2) the beam widths
are large ({\sc fwhm}~$\simeq 15\arcsecs$ in the case of {\sc scuba}), 
and that c) the two sources are clearly extended at 850-$\mu$m. \\
\indent The RG is marginally resolved at 350-$\mu$m, and exhibit the same general shape as
seen at 850-$\mu$m, thus apparently confirming the claim by
Stevens et al.\ (2003) that the (850-$\mu$m) dust emission in this object (and its 
brightest companion) is extended on tens
of kpc scales. This seems to be backed up by the detection of a weak ($\sim 4\sigma$,
$S_{350\mu m}=70\pm 17$\,mJy)
source only $\sim 19\arcsecs$ north-east of the RG, and thus within the north-east extended
850-$\mu$m emission from the RG.\\
\indent A third much fainter {\sc scuba} source, $\sim 35\arcsecs$ west of the RG and 
within the {\sc sharc-ii} map, is not detected at 350-$\mu$m, and an upper flux limit of 
$S_{350\mu m} \ls 28$\,mJy is derived. 

\begin{figure}[h]
\resizebox{8.7cm}{!}
{\includegraphics[]{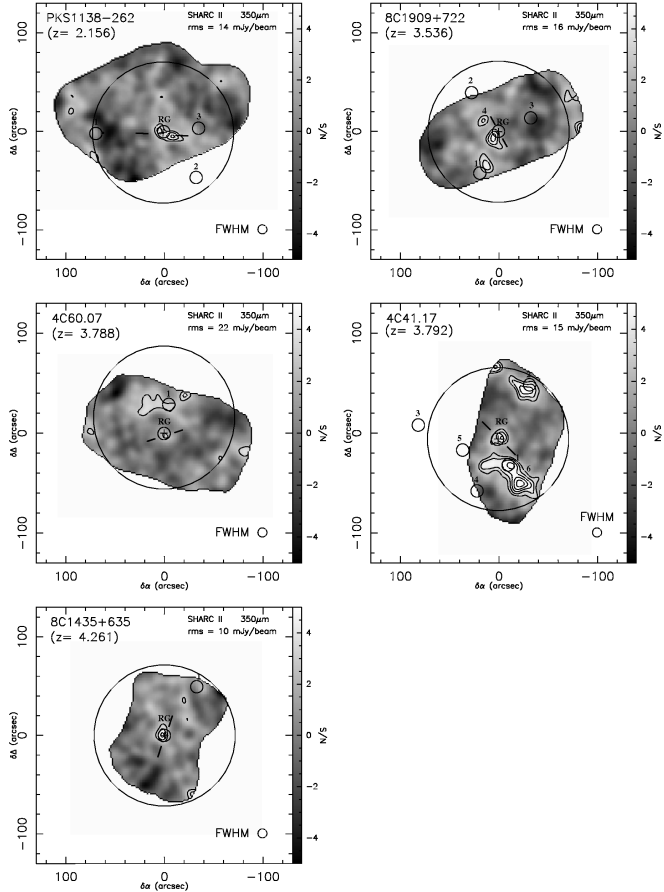}}
\caption{{\small {\sc sharc-ii} 350-$\mu$m signal-to-noise maps of the five radio galaxies observed. Contours are
at $3, 4, 5, 6\sigma$, where $\sigma$ is the rms noise in each map, see the individual panels.
The {\sc sharc-ii} beam at 350-$\mu$m ({\sc fwhm}~$= 9\arcsecs$) is shown in the lower right
corner of each panel. The crosses mark the position of the radio core. The open circles and the
associated labels indicate the 850-$\mu$m sources detected with {\sc scuba} by 
Stevens et al.\ (2003) and Ivison et al.\ (2000). The tick marks around the central radio galaxy indicate
the alignment of the large-scale radio jets, and the large circles outline the extend of the
{\sc scuba} maps of Stevens et al.\ (2003).}}
\label{figure:maps}
\end{figure}

\paragraph{4C\,60.07}
This source received by far the least amount of integration time of the 5 HzRGs
observed with {\sc sharc-ii}, and as a result the source is only just detected
at the $\sim 3\sigma$ level ($S_{350\mu m}=59\pm 22$\,mJy). This source
was one of the strongest sources detected at 850-$\mu$m by Stevens et al.\ and 
even showed some, albeit marginal, evidence of being extended. 
A faint 850-$\mu$m source found $\sim 30\arcsecs$ north of the RG 
coincides with extended $3\sigma$ emission in the {\sc sharc-ii} map. However,
given its small 850-$\mu$m flux ($S_{350\mu m}=5.9\pm 1.3$\,mJy), and the
somewhat larger noise level in that part of the {\sc sharc-ii} map, we do not
consider this source robustly detected at 350-$\mu$m. We derive a $3\sigma$ upper
flux limit of $S_{350\mu m} \ls 91$\,mJy

\paragraph{4C\,41.17}
With 5 sources detected at $\gs 5\sigma$ significance, this is by far
the richest field observed at 350-$\mu$m. All three {\sc scuba} sources
detected by Stevens et al., including the RG which even appears to be
slightly extended, are detected at 350-$\mu$m, 
and their flux densities are listed in Table 1. Ivison et al.\ (2000) reported an
additional three faint {\sc scuba} sources. However, only one of those sources lies
within the {\sc sharc-ii} map and is not detected to a $3\sigma$ limit
of $S_{350\mu m}\ls 67$\,mJy. \\
\indent The most significant ($\sim 8\sigma$) 350-$\mu$m source lies $\sim 53\arcsecs$ south-west of the RG,
and although it is not detected at 850-$\mu$m it appears to be associated with the very
bright {\sc scuba} (and {\sc sharc-ii}) source $\sim 30\arcsecs$ north-east of it. 
The two 350-$\mu$m sources make up a large, extended structure aligned along the jet-direction. 
The reality of this structure, and of the most south-western source in particular, will
have to be confirmed by independent submm observations.
Finally, a $\sim 5\sigma$ source is detected at 350-$\mu$m, about 68\arcsecs north of the RG.
Unfortunately, the source lies outside the {\sc scuba} map and thus cannot be confirmed
at 850-$\mu$m. Given that this source lies right at the map edge where the noise is higher, 
we suspect that it may be spurious.\\
\indent Although, we cannot exclude that some of the {\sc sharc-ii} sources (in particular
the ones closest to the map edges) are spurious, the confirmation of the 
three {\sc scuba} sources reported by Stevens et al.\ (2003), supports their 
conclusions that the environment of 4C\,41.17 has an overdensity of submm sources.

\paragraph{8C\,1435$+$635}
Despite being the faintest radio galaxy at 850-$\mu$m,
the central radio galaxy is clearly detected ($\gs 5\sigma$) in the {\sc sharc-ii} maps
($S_{350\mu m} = 51\pm 9$\,mJy). Similar to Stevens et al., who failed to resolve the galaxy at 850-$\mu$m,
there is no evidence of extended emission at 350-$\mu$m either,
suggesting that this source may be relatively compact.
The only other {\sc scuba} source detected in the 8C\,1435$+$635
field lies about $60\arcsecs$ north-west of the RG and coincides with faint
$2\sigma$ emission at 350-$\mu$m. We derive an upper flux limit
of $S_{350\mu m} \ls 68$\,mJy.

\begin{figure}
\resizebox{\hsize}{!}
{\includegraphics[width=1.0\hsize,angle=0]{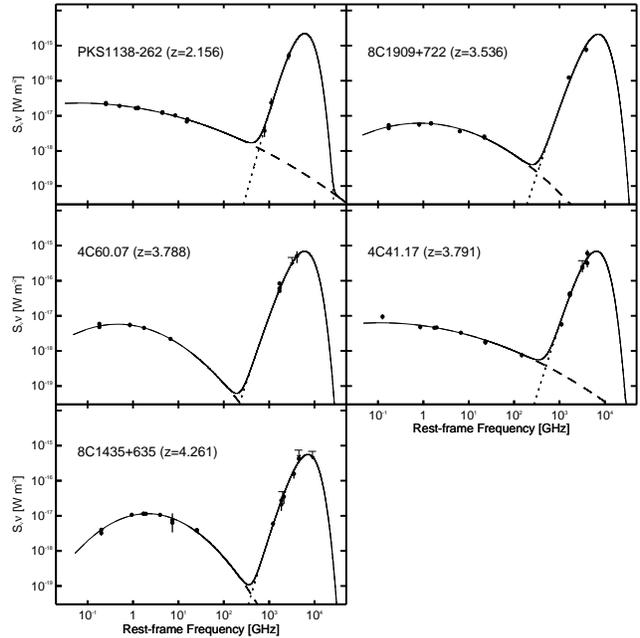}}
\caption{{\small Radio-FIR/submm spectral energy distributions of the 5 HzRGs observed with
{\sc sharc-ii}. The dust temperature and spectral index were allowed to vary freely in the fitting process,
except in the case of 8C\,1909$+$722 where $\beta$ was fixed at 1.0 due to the small
number of data point. The radio part of the spectrum was fitted with a parabola (dashed curves). The
combined radio to FIR/submm SEDs are shown as solid curves.}}
\label{figure:seds}
\end{figure}
\begin{table*}[t]
\begin{center}
\caption{Observed 350-, 450-, 850- and 1200-$\mu$m fluxes and derived properties of the 5 HzRGs and their associated submm
sources (which fall within the {\sc sharc-ii} maps). The numbering follows that of
Stevens et al.\ (2003) and Ivison et al.\ (2000).}
\label{tlab}
\begin{tabular}{lcccccccccc}\hline
Source           & $z$    & $S^{a}_{350\mu m}$  & $S_{450\mu m}^b$  & $S_{850\mu m}^c$     & $S_{1200\mu m}^d$ & $T_d$     & $\beta$      & $L_{\mbox{\tiny{FIR}}}$    &  $M_{d}$           \\ 
                 &        & [mJy]               &  [mJy]            & [mJy]                & [mJy]             & [K]       &              & [$\times 10^{13} \Lsolar$] &  [$\times 10^8 \Msolar$]       \\
\hline
PKS\,1138$-$262a & 2.156  & $61\pm 15$          & . . .             & $3.7\pm 2.4$         & $1.5\pm 0.5^e$    & $49\pm 6$ & $1.9\pm 0.2$ &  $2.1\pm 0.5$              & $2\pm 1$       \\
PKS\,1138$-$262b & 2.156  & $92\pm 18$          & . . .             & $3.8\pm 2.4$         & . . .             & . . .     & . . .        & . . .                      & . . .              \\
          1.     & . . .  & $\le95$             & . . .             & $7.8\pm 2.2$         & . . .             & . . .     & . . .        & . . .                      & . . .              \\
          3.     & . . .  & $\le68$             & . . .             & $2.2\pm 1.4$         & . . .             & . . .     & . . .        &  . . .                     & . . .              \\
8C\,1909$+$722   & 3.536  & $90\pm 15$          & . . .             & $34.9\pm 3.0$        & . . .             & $69\pm 3$ & $1.0^e$      & $5.7\pm 0.9$               & $13\pm 2$          \\
          1.     & . . .  & $148\pm 32$         & . . .             & $23.0\pm 2.5$        & . . .             & . . .     & . . .        & . . .                      & . . .              \\
          3.     & . . .  & $\le 28$            & . . .             & $4.3\pm 2.1$         & . . .             & . . .     & . . .        & . . .                      & . . .              \\
          4.     & . . .  & $70\pm 17$          &                   & . . .                & . . .             & . . .     & . . .        &  . . .                     & . . .              \\
4C\,60.07        & 3.788  & $59\pm 22$          & $69\pm 23$        & $23.8\pm 3.5$        & . . .             & $58\pm 8$ & $1.0\pm 0.2$ & $2.9\pm 1.1$               & $15\pm 6$          \\
          1.     & . . .  & $\le91$             & . . .             & $6.3\pm 2.1$         & . . .             & . . .     & . . .        & . . .                      & . . .              \\
4C\,41.17        & 3.791  & $71\pm 17$          & $35.3\pm 9.3$     & $12.0\pm 2.3$        & $2.7\pm 0.3$      & $60\pm 3$ & $1.3\pm 0.1$ & $4.1\pm 0.9$               & $8\pm 2$          \\
          1.     & . . .  & $108\pm 17$         & $34.1\pm 9.3$     & $21.2\pm 2.9$        & . . .             & . . .     & . . .        & . . .                      & . . .              \\
          2.     & . . .  & $184\pm 34$         & . . .             & $6.2\pm 1.9$         & . . .             & . . .     & . . .        & . . .                      & . . .              \\
          4.     & . . .  & $\le 67$            & $\le 26$          & $2.8\pm 0.8$         & . . .             & . . .     & . . .        & . . .                      & . . .              \\
          6.     & . . .  & $197\pm 24$         & . . .             & $\le 5$              & . . .             & . . .     & . . .        & . . .                      & . . .              \\
          7.     & . . .  & $181\pm 40$         & . . .             & . . .                & . . .             & . . .     & . . .        & . . .                      & . . .              \\
8C\,1435$+$635   & 4.261  & $51\pm 9$           & $23.6\pm 6.4$     & $6.0\pm 2.1$         & . . .             & $64\pm 5$ & $1.4\pm 0.1$ & $3.8\pm 0.7$               & $4\pm 1$          \\
          1.     & . . .  & $\le68$             & . . .             & $3.9\pm 1.8$         & . . .             & . . .     & . . .        & . . .                      & . . .              \\
\hline
\end{tabular}
\end{center}
{\tiny $^a$ The total flux density within a $r=15\arcsecs$ aperture. The flux errors include the $\sim 15$ per-cent calibration uncertainty; 
$^b$ 450-$\mu$m flux densities are from Archibald et al.\ (2002) and Ivison et al.\ (2000); $^c$ 850-$\mu$m flux densities are from Stevens 
et al.\ (2003), except for PKS\,1138$-$262a and b which are our own measurements, and 4C\,41.17-4 which is from Ivison et al.\ (2000);
$^d$ 1200-$\mu$m flux densities are from Chini \& Kr\"{u}gel (1994) and Carilli et al.\ (2001); $^e$ Fixed value.\\}
\end{table*}

\section{Spectral Energy Distributions, FIR luminosities and Dust Masses}
For each of the central radio galaxies, an optically thin grey-body spectrum of the form
$S_{\nu} \propto \nu^{3+\beta} ( e^{h\nu/kT_d} -1 )^{-1}$, 
was fitted using all FIR/submm data points 
available in the literature. In all cases but 8C\,1909$+$722 more than one independent 850-$\mu$m
{\sc scuba} measurement existed (Ivison et al.\ 1995, 1998;
Hughes et al.\ 1997; Archibald et al.\ 2001, Reuland et al.\ 2004), all of which were
included in the fitting-process, and in a few
cases even 1200-, 450- and/or 350-$\mu$m observations had been made.
In the case of 8C\,1909$+$722 where the FIR/submm SED has been sampled
at only two separate frequencies we allowed only $T_d$ to vary in the fit
for fixed values of $\beta=1.0, 1.5$ and $2.0$. The best fit was found 
for $\beta=1.0$.
The resulting dust SEDs are shown as dotted curves in Figure \ref{figure:seds}, and
the fitted $T_d$ and $\beta$ values are listed in Table 1.
The radio part of the spectrum was fitted with a parabolic function (dashed curves),
and the combined FIR/submm-radio SED is shown as solid curves in Figure \ref{figure:seds}.\\
\indent Due to the fact that our 350-$\mu$m observations probe the SEDs near their peaks at 
$\sim 100$-$\mu$m rest-frame wavelengths, we are able to constrain 
dust temperatures and FIR luminosities much better than previously. 
In general, the derived dust temperatures are rather high (lowest value is $T_d=49$\,K),
compared to the median dust temperatures of submm-selected galaxies ($T_d\sim 36$\,K --
Chapman et al.\ 2004). While HzRGs, in contrast to SMGs where the
AGN rarely plays a dominant role (Alexander et al.\ 2003), 
harbour powerful AGN, capable of heating the dust in the nuclear
regions to high temperatures, it is difficult to envisage them doing so
on scales of $\gs 10$\,kpc as suggested by the extended 350-$\mu$m emission.
The spectral emissivity index varies between $\beta\sim 1-2$, a similar
range to what is observed in SMGs. We note that for objects at $z\sim 3$,
$\beta$ is much better constrained by 850- and/or 1200-$\mu$m
observations which probe further down the Rayleigh-Jeans tail than 350-$\mu$m.\\
\indent FIR-luminosities obtained by integrating the FIR/submm SED over the wavelengths
range $200-1000\,\mu$m range are listed in Table 1, along with
dust masses derived from the observed 350-$\mu$m fluxes using 
a mass absorption coefficient
of $\kappa(\nu_r)=0.11(\nu_r/352\mbox{GHz})^{\beta}$ in units of
m$^2$\,kg$^{-1}$. Assuming that only half of the FIR-luminosity in these objects
is due to star formation, we find that on average they form stars at a rate 
of $\sim 1500\,\Msolar\,$yr$^{-1}$.
The large dust masses suggest that star formation must have been
going on for some time in these systems.
That such large star formation rates can be sustained for a long enough time to
form a giant elliptical ($\sim 1$\,Gyr) is shown in the cases of 4C\,60.07 and 4C\,41.17   
where the detection of CO has revealed large amounts of molecular gas
($\sim 10^{11}\,\Msolar$ -- Papadopoulos et al.\ 2000; Greve et al.\ 2004; De Breuck et al.\ 2005).

\section{Concluding Remarks}
To conclude, our SHARC-II maps confirm the over-densities of dusty galaxies 
detected in HzRG fields by SCUBA. Furthermore, the superior spatial 
resolution of these data allow us to confirm that dust emission associated 
with both the HzRGs and companions is extended on galaxy-wide scales. Our 
results suggest that mergers play a key role in the formation of these 
galaxies.

\end{document}